\documentclass[prc,aps,showpacs,groupedaddress,floatfix,preprint]{revtex4}  
\usepackage{epsfig}
\usepackage{graphicx}

\begin{document}
\title{The Onset of Nuclear Structure Effects in Near-Barrier Elastic Scattering of Weakly-Bound Nuclei:
$^6$He and $^6$Li Compared.}
\author{Y. Kucuk, I. Boztosun,}
\affiliation{Department of Physics, Erciyes University, Kayseri, Turkey}
\author{N. Keeley}
\affiliation{The Andrzej So\l tan Institute for Nuclear Studies, Department of Nuclear Reactions,
ul.\ Ho\.za 69, 00-681 Warsaw, Poland}

\begin{abstract}

The elastic scattering of the halo nucleus $^6$He from heavy targets at incident
energies near the Coulomb barrier displays a marked
deviation from the standard Fresnel-type diffraction behavior. This deviation
is due to the strong Coulomb dipole breakup coupling produced by the Coulomb
field of the heavy target, a specific feature of the nuclear structure of
$^6$He. We have performed Continuum Discretized Coupled Channels calculations
for the elastic scattering of $^{6}$He and $^6$Li from
$^{58}$Ni, $^{120}$Sn, $^{144}$Sm, $^{181}$Ta and $^{208}$Pb targets in order to determine the
range of $Z_{\mathrm T}$ where this nuclear-structure specific coupling effect becomes manifest.
We find that the strong Coulomb dipole breakup coupling effect
is only clearly experimentally distinguishable for targets of $Z_{\mathrm T} \approx 80$.
\end{abstract}

\maketitle
The interaction of two composite nuclei may lead to strong absorption in which the
effects of coupling to non-elastic channels are dominant \cite{Satchler}.
When strong absorption occurs, the scattering is primarily diffractive
in nature and the elastic scattering cross section angular distributions
will be of one of two types, characteristic of Fresnel or Fraunhofer
diffraction, depending on the incident energy. Fresnel-type diffraction is
observed at energies close to the Coulomb barrier when the Coulomb field
acts like a diverging lens. As the incident energy is increased, the behavior
of the angular distribution transforms from Fresnel- to Fraunhofer-type
scattering where the Coulomb field is no longer effective as a diverging
lens and interference between waves diffracted around opposite edges of the
targets occurs, leading to the characteristic oscillatory behavior.

While stable nuclei usually exhibit one or other of
these classical diffraction patterns in their elastic scattering
angular distributions, the elastic scattering of
the $^6$He halo nucleus from heavy targets at near-barrier energies shows a
strong deviation from the standard diffraction behavior.
At these energies we would expect the elastic scattering to display
the characteristic Fresnel-type diffraction pattern. However, a different
structure is observed, in that the usual Coulomb
rainbow peak is completely absent \cite{Rus03,Mor07}.
The $^6$Li nucleus shows a similar anomalous scattering for heavy targets
at near-barrier incident energies  but
it is much weaker and considerably more difficult to observe experimentally,
being a reduction of the Coulomb rainbow peak rather than a complete
absence as for $^6$He, requiring very precise measurement of the elastic scattering
angular distributions \cite{Kee94}.

A similar deviation from the classical Fresnel diffraction pattern was initially
observed experimentally in the elastic scattering of $^{18}$O + $^{184}$W \cite{Tho77},
and was interpreted as arising
from the effect of strong Coulomb excitation of the first $2^+$ state
in the $^{184}$W target.
Strong Coulomb coupling effects are also responsible for the effect
seen in the near-barrier elastic scattering of $^6$He from heavy targets.
When the atomic number of the target nucleus ($Z_{\mathrm T})$ is large, the
breakup of the weakly bound projectile is dominated by the
Coulomb field. It is the large Coulomb dipole ($E1$) breakup probability of $^6$He
and the strong coupling of this process to the elastic scattering
that causes the deviation for a $^6$He projectile from the classical
diffraction pattern. The elastic scattering of $^6$He from the medium-mass
$^{64}$Zn target does not show this effect \cite{Mor07,Dip04}, and
appears similar to that for $^6$Li from similar mass targets, presumably due to the
reduced importance of the Coulomb breakup. For $^6$Li, the similar effect on the
elastic scattering is caused by the virtual \emph{quadrupole} ($E2$) breakup coupling ($E1$ breakup is
not allowed for the $^6$Li $\rightarrow$ $\alpha + d$ process) and
is consequently much weaker than for $^6$He (with both $E1$ and $E2$ breakup allowed) and only apparent in precise
measurements for heavy targets like $^{208}$Pb at near-barrier energies.
The elastic scattering of $^6$Li therefore provides a good benchmark for comparison
with $^6$He elastic scattering.

It is not possible at present to easily control the beam energy of
radioactive nuclei and thus optimize the experimental visibility of any
interesting features that may arise due to the particular internal
structure properties of these nuclei. For
example, one feature of halo nuclei is the possibility of
low-lying dipole strength, and this characteristic has been demonstrated
experimentally and theoretically in the scattering of $^6$He from $^{208}$Pb.
The change observed in the elastic scattering is an interference
between nuclear and Coulomb contributions that is highly dependent on
the charge of the target nucleus and the beam energy relative to the
Coulomb barrier.  While it is not possible yet to predict all of the
other types of behavior that might occur in exotic nuclei, exploring the
virtual dipole effect as a function of bombarding energy and nuclear
target charge theoretically for $^6$He scattering can show the regime where
one should look generally for these new effects and where the elastic
scattering is sensitive to the details of the projectile nuclear structure.

In this note, we investigate over what range of $Z_{\mathrm T }$ the
Coulomb dipole breakup virtual coupling effect is sufficiently important
that the $^6$He elastic scattering shows a measurable difference from
the analogous $^6$Li scattering and is therefore sensitive to its
specific nuclear structure properties.
For this purpose, we have calculated the elastic
scattering of $^{6}$He and $^6$Li by different nuclei from $^{58}$Ni to
$^{208}$Pb at energies near the Coulomb barrier using the
Continuum Discretized Coupled Channels (CDCC) method. As the strong
coupling effect is linked to the specific nature of the $^6$He structure
it is hoped that this study will prove useful in planning future radioactive
beam experiments by helping to pinpoint the target and incident energy ranges
where such structure-dependent effects are most clearly manifest.

We performed CDCC calculations for ten different systems,
$^{6}$He and $^6$Li + $^{58}$Ni, $^{120}$Sn, $^{144}$Sm, $^{181}$Ta and $^{208}$Pb
in order to find a critical $Z_{\mathrm T}$ value where the
$^6$He elastic scattering is
measurably different from that for $^6$Li. In order to remove trivial effects
due to the difference in charge between $^6$Li and $^6$He, calculations were compared
for the same centre of mass energy relative to the Coulomb barrier, $E_{\mathrm{c.m.}} -
V_{\mathrm B}$, where the Coulomb barrier height $V_{\mathrm B}$ was calculated
according to the relation \cite{Hod78}:
\begin{equation}
V_{\mathrm B}  =  \frac{Z_{\mathrm{P}} Z_{\mathrm{T}} e^2}{R_{\mathrm{P}} + R_{\mathrm{T}}}
\end{equation}
where $R = 1.16 \; \; {\mathrm A}^{1/3} + 1.2$. While this relation overestimates the
Coulomb barrier due to its neglect of the nuclear potential it should be adequate for our
purposes. For each target, calculations were performed at two energies, corresponding to
values of $E_{\mathrm{c.m.}} - V_{\mathrm B}$ of $1.005$ and $5.534$ MeV, equating to
incident laboratory frame energies of 11.0 and 16.0 MeV for the $^6$He + $^{58}$Ni system.

Although $^{6}$He has a three-body $\alpha+n+n$ structure, assuming
an $\alpha$ + $^2n$ cluster structure can give physically meaningful
results as the three-body wave function of the $^{6}$He ground state
has a large di-neutron ($^{2}n$) component, which dominates the tail
of the wave function \cite{Zhukov,Rus03}. Thus, while the breakup of
$^6$He is best described by four-body models
\cite{Mat04,Mat04a,Mat06,fb1,fb2}, their numerically demanding
nature, combined with the lack of a generally available code able to
implement such calculations make the use of standard three-body CDCC
calculations attractive in a study of this kind. Therefore, CDCC
calculations were performed using the modified two-body di-neutron
model of $^{6}$He proposed by Moro \emph{et al.} \cite{Mor07}, where
the binding energy of the di-neutron in the ground state is
increased to 1.6 MeV to give a wave function that well matches that
of more physically sophisticated three-body models. This model
describes very well the elastic scattering of $^6$He for several
targets covering the mass range studied here and gives a coupling
effect similar to that of four-body CDCC calculations \cite{Mor07}.
It is therefore adequate for our purposes in providing a good
description of the elastic scattering of $^6$He, although it is not
claimed to provide an accurate picture of the breakup cross section
itself, merely its coupling effect on the elastic scattering. To
calculate the interaction potentials the single-folding technique
\cite{Buck} was used and the necessary $\alpha$ + target, $^2n$ +
target optical potential parameters were taken from Refs.\
\cite{Avrigeanu} and \cite{Perey}, respectively, the latter being a
global deuteron potential as $^2n$ scattering potentials are
obviously not available. The $\alpha$ + $^{2}n$ binding potential
was of Woods-Saxon form with parameters $R=1.9$ fm and $a=0.25$ fm
\cite{Rusek2}.

The $^{6}$He $\alpha$ + $^{2}n$ continuum was discretized into bins
of widths $\triangle k$ = 0.1 fm$^{-1}$ and up to a maximum
excitation energy of $\epsilon=7.7$ MeV in $\alpha$ + $^2n$ relative
momentum $(k)$ space. The maximum value of $k$ was chosen in each
case so as to ensure convergence of the results, i.e.\ it was
checked that adding an additional bin did not affect the result of
the calculation. All non-resonant cluster states corresponding to
$\alpha$ + $^2n$ relative angular momenta $L=0,1,2,3$ were included
as well as the 1.8 MeV 2$^{+}$ resonant state. The coupled equations
were integrated up to $R=80$ fm and used 200 partial waves for the
projectile-target relative motion.

The $^6$Li calculations were similar to those described in Ref.\ \cite{Bec07}.
Again, the maximum value of $k$ was chosen to ensure convergence. The $\alpha$
+ target and $d$ + target potentials were also taken from Refs.\ \cite{Avrigeanu}
and \cite{Perey}, respectively.
All calculations were performed using the code Fresco \cite{Thompson}.

\begin{figure}
\psfig{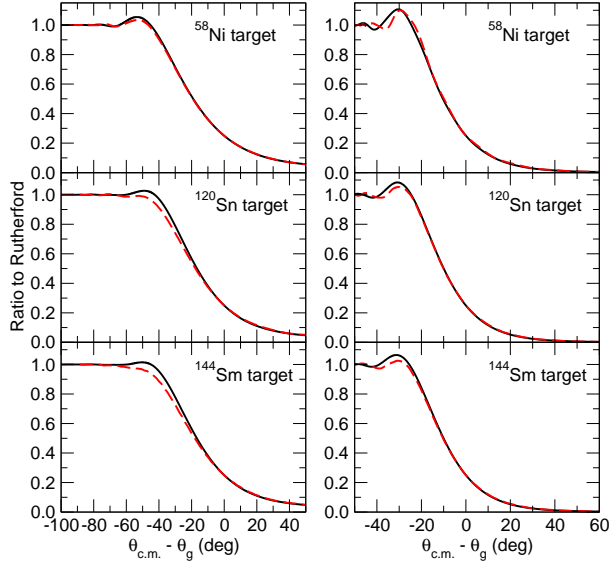}
\caption{(Color online) Angular distributions of differential cross section (ratio to
Rutherford cross section) for $^{6}$He (dashed curves) and $^6$Li (solid curves) + $^{58}$Ni,
$^{120}$Sn, and $^{144}$Sm elastic scattering. The left-hand panels are for $E_{\mathrm{c.m.}}
- V_{\mathrm B} = 1.005$ MeV and the right-hand panels for $E_{\mathrm{c.m.}}
- V_{\mathrm B} = 5.534$ MeV. \label{fig1}}
\end{figure}

\begin{figure}
\psfig{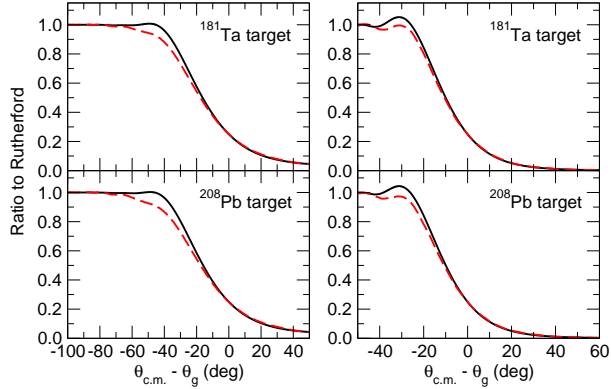}
\caption {(Color online) Angular distributions of differential cross section (ratio to
Rutherford cross section) for $^{6}$He (dashed curves) and $^6$Li (solid curves)
+ $^{181}$Ta and $^{208}$Pb elastic scattering. The left-hand panels are for $E_{\mathrm{c.m.}}
- V_{\mathrm B} = 1.005$ MeV and the right-hand panels for $E_{\mathrm{c.m.}}
- V_{\mathrm B} = 5.534$ MeV. \label{fig2}}
\end{figure}

The results of the calculations are presented in Figs.\ \ref{fig1} and \ref{fig2},
those for $^6$He being denoted by the dashed curves and those for $^6$Li
by the solid curves. To emphasize the angular region around the Coulomb rainbow
the cross section scales (expressed as a ratio to the Rutherford cross section)
are linear. To remove any residual ``geometric'' differences the angular distributions
are plotted as a function of $\theta_{\mathrm{c.m.}} - \theta_{\mathrm g}$, where
$\theta_{\mathrm g}$ is the grazing angle defined by the ``quarter-point recipe''.

We immediately see that for a $^{58}$Ni target ($Z_{\mathrm T} = 28$)
the calculated $^6$Li and $^6$He angular distributions
are absolutely identical when plotted in this fashion; measurements of the elastic
scattering from targets in this mass region are clearly not sensitive to the details
of the nuclear structure of the projectile. For a $^{120}$Sn target, while the calculated
angular distributions are slightly different at $E_{\mathrm{c.m.}} - V_{\mathrm B} =
1.005$ MeV the difference is too small to be measurable; at $E_{\mathrm{c.m.}} - V_{\mathrm B}
= 5.534$ MeV the $^6$He and $^6$Li angular distributions are again identical.

The magnitude of the structure-specific coupling effect for $^6$He elastic scattering
of course increases with increasing target charge, it being a consequence of strong
Coulomb dipole coupling; thus for the $^{144}$Sm target
($Z_{\mathrm T} = 62$) a complete lack of a Coulomb rainbow is clearly observed
for $E_{\mathrm{c.m.}} - V_{\mathrm B} = 1.005$ MeV, although any reasonable measurement
would still be unable to detect any difference from the corresponding $^6$Li angular distribution.
The $^{144}$Sm target also provides a good example of the dependence of the coupling
effect on incident energy, as the calculated angular distribution for for $E_{\mathrm{c.m.}}
- V_{\mathrm B} = 5.534$ MeV is virtually identical to the $^6$Li one. For a given target,
as the incident energy is increased the coupling effect weakens and the Coulomb rainbow
gradually manifests itself. This is a well-known general feature of Fresnel-type scattering
for heavy ions; for $^6$He scattering from a heavy target the effect is somewhat different
as the Coulomb breakup coupling dominates at energies just above the Coulomb barrier
to such an extent that the Coulomb rainbow is not merely absent but completely
effaced.

In Fig.\ \ref{fig2} the angular distributions for $^{181}$Ta ($Z_{\mathrm T} = 73$)
and $^{208}$Pb ($Z_{\mathrm T} = 82$) targets show clear differences between $^6$He
and $^6$Li at both values of $E_{\mathrm{c.m.}} - V_{\mathrm B}$. However, for the
$^{181}$Ta target the difference at $E_{\mathrm{c.m.}} - V_{\mathrm B} = 1.005$ MeV would
be barely detectable in a measurement to a precision of $\pm 1$ \% for the the
$^6$Li elastic scattering and $\pm 2$ \% for the $^6$He measurement (both achievable
in a reasonable time scale with currently available beam intensities and detector arrays).
The use of a Ta target is largely hypothetical in any case, as all the stable isotopes
of this element have very low-lying excited states that make the measurement of pure
elastic scattering impossible, even with stable beams. This problem also occurs for
the other elements in the $Z = 70$ region, ruling out their practical use as targets
in this type of study; we included a  $^{181}$Ta target in our study for the sake
of completeness to check whether a (hypothetical) ideal target with a charge of
around 70 would be sufficient to enable clear experimental separation of projectile
structure-specific coupling effects in the elastic scattering.

\begin{figure}
\psfig{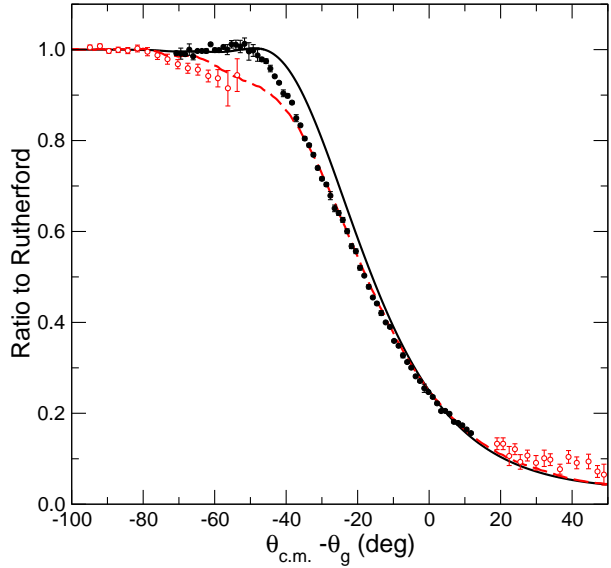} \caption{(Color online)
Experimental angular distributions of the differential cross section
(ratio to Rutherford cross section) for $^{6}$He (unfilled circles)
and $^6$Li (filled circles) elastic scattering from $^{208}$Pb at
incident energies of 22.0 MeV \cite{San08} and 33.0 MeV
\cite{Kee94}, respectively. Solid and dashed curves denote CDCC
calculations for $^6$Li and $^6$He projectiles,
respectively.\label{fig3}}
\end{figure}

With a $^{208}$Pb target we finally see a clearly measurable difference between the
$^6$He and $^6$Li elastic scattering angular distributions for $E_{\mathrm{c.m.}} -
V_{\mathrm B} = 1.005$ MeV; at $E_{\mathrm{c.m.}} - V_{\mathrm B} = 5.534$ MeV
the difference would just be detectable for measurements with a precision of
$\pm 1$ \% and $\pm 2$ \% for $^6$Li and $^6$He, respectively. Values of
$E_{\mathrm{c.m.}} - V_{\mathrm B} = 1.005$ MeV correspond to incident $^6$Li
and $^6$He energies of 33.06 MeV and 22.38 MeV, respectively for a $^{208}$Pb
target. Measured elastic scattering angular distributions for $^6$Li and $^6$He
+ $^{208}$Pb are available in the literature for incident energies of 33.0 MeV
\cite{Kee94} and 22.0 MeV \cite{San08}, enabling us to test the reliability of
our calculations and the conclusions to be drawn therefrom. We plot them
as a function of $\theta_{\mathrm{c.m.}} - \theta_{\mathrm g}$ in Fig.\ \ref{fig3},
together with the relevant CDCC calculations.
Not only do they confirm the results of our calculations, but also the practicability
of measuring the elastic scattering to sufficient precision to observe the predicted
effect. The agreement between calculations and data is not perfect due to the use of
global optical potentials as input in order to have a consistent set of results for
several targets; slight adjustment of the potential well depths or the
use of fitted potentials would enable perfect fits to be obtained. However,
in the context of this work only qualitative agreement is required.

In summary, it has been shown by means of CDCC calculations that the large Coulomb dipole coupling
effect observed in the elastic scattering of $^6$He from $^{197}$Au and $^{208}$Pb targets
at energies close to the Coulomb barrier \cite{Mor07} is only clearly evident, in the sense
that the angular distribution is unambiguously experimentally distinguishable from
that for $^6$Li, for targets with $\mathrm{Z}_T \approx 80$. Furthermore, the incident energy
must be close to the top
of the nominal Coulomb barrier, that is to say a few MeV above the experimentally
determined barrier (in the sense of the energy at which the measured elastic scattering cross section becomes
equal to that for Rutherford scattering over the entire angular range).

The calculations presented here are specific to $^6$He scattering.
However, it is now well established that $^6$He has a strong
low-lying electric dipole strength in the $\alpha + n + n$
continuum, see e.g.\ \cite{Aum99}, and that coupling to this
strength is responsible for the characteristic appearance of the
elastic scattering of $^6$He from heavy targets \cite{Rus03,Mor07}.
Low-lying continuum dipole strength is a property that is
shared, or thought to be shared, with several other weakly-bound
light radioactive nuclei, e.g. $^{11}$Li and $^{11}$Be. While
detailed comparison with models of the nuclei in question remains
difficult due to the many-body nature of the problem (although
progress is being made in this direction, see e.g.\
\cite{Mat04,Mat04a,Mat06,fb1,fb2,Sum06,Sum06a}) it should be
possible to make qualitative conclusions concerning the relative
strengths of these couplings by a comparison of the relevant
near-barrier elastic scattering measurements. Extrapolation of the
calculations presented here leads to the conclusion  that future
experiments to measure the elastic scattering of such nuclei should
concentrate on heavy ($\mathrm{Z}_T \approx 80$) targets ---
preferably $^{208}$Pb or similar --- at energies a few MeV above the
Coulomb barrier for the systems concerned in order to maximize the
structure dependence of the coupling effects.

\section*{Acknowledgments}
This project is supported by the Turkish Science and Research
Council (T\"{U}B\.{I}TAK) with Grant No:107T824 and the Turkish
Academy of Sciences (T\"{U}BA-GEB\.{I}P). Y. Kucuk would also like
to thank Prof. K. Rusek for the hospitality during her stay in
Warsaw.

\end{document}